\documentclass[aps,pre,onecolumn,showpacs,showkeys,a4paper]{revtex4}
\input epsf  
\usepackage{graphicx} 
\usepackage{amsmath}
\usepackage{amssymb}
\usepackage{amscd}

\begin{document} 

 \title{An Exact Formula  for the   Statistics  of the Current
  in the  TASEP with Open Boundaries}
\author{Alexandre Lazarescu and Kirone Mallick}
  \affiliation{Institut  de Physique Th\'eorique, C. E. A.  Saclay,
 91191 Gif-sur-Yvette Cedex, France}
\pacs{05-40.-a; 05-60.-k; 02.50.Ga}
 \keywords{Non-equilibrium statistical physics; current fluctuations; large deviations;
 ASEP; open boundaries.}
 \begin{abstract} 
     We study the totally asymmetric exclusion process (TASEP)  on a finite one-dimensional
 lattice  with open boundaries, i.e.,  in contact with two reservoirs
 at different potentials. The total (time-integrated)  current   through the system
 is a random variable that  scales  linearly  with time  in the long time limit.
 We give a parametric  representation  for the generating function of the cumulants of
 the current, which is related to the large deviation function by Laplace transform.
 This formula is valid for all system sizes and for all values
 of the  boundary coupling parameters. 
  \end{abstract}

\maketitle 

\section{Introduction}

 The asymmetric simple exclusion process (ASEP) is one of the fundamental
 models in  non-equilibrium statistical mechanics.
 The ASEP involves  particles that perform  asymmetric jumps on a discrete
 lattice under the exclusion constraint: two particles cannot occupy
 the same  site at the same time. This  very simple and
 minimal system  appears  as a building block in more
 realistic descriptions  for low-dimensional  transport with constraints.
 Such phenomena   occur
 in various contexts and scales ranging from micrometric cellular motors
 to   traffic networks \cite{PaulK,Zia}.
 The  remarkable  properties of the ASEP and its numerous variants
 have stimulated
 hundreds of studies during the last two decades.  The ramifications
 of this mundane-looking model through non-equilibrium statistical mechanics,
 combinatorics, probability, random matrices and representation
 theory  are tremendous
  \cite{MartinReview,DerridaRep,FerrariPatrick,PaulK,Sasamoto,Schutz,Spohn}.

 Generically, a system out of equilibrium  carries at least one 
 non-vanishing current in its steady  state  (this is related
 to the breaking of detailed balance). 
 Such  currents  can be considered
 as archetypal observables  for non-equilibrium behaviour \cite{Zia}.
  Classifying
 the different independent stationary   currents in a given  system,  
  identifying some generic features
 and relations  obeyed by the distributions of these currents and calculating
 their statistical properties as functions of the control parameters,  are some 
 of the important tasks  in non-equilibrium statistical physics. Current
 fluctuations  are usually non-Gaussian:
 their characteristics  can be quantified  by the  moments of the current
 (mean-value, variance, skewness,
 kurtosis...) or by a  large-deviation function that measures the
 probability for  the current to assume a   non-typical value.
  There is  growing  evidence that  large-deviation functions  play a crucial
 role in non-equilibrium statistical physics, akin to that of thermodynamic
 potentials at equilibrium \cite{DerrReview,Touchette}.

   Currents transport information from one  part of the system
 to another. In particular,  in 
  a  system far from  equilibrium, boundary conditions can drastically alter
 the behaviour of the bulk  even if interactions are short-ranged
 (in contrast with the generic case at  equilibrium).
 For example, as  was recognized in the earliest studies \cite{Krug,Janowski},
  phase-transitions
 can be induced by   the boundaries or by  a localized  alteration
 of the dynamical rules. It is therefore  crucial
 to specify the boundary conditions (e.g. periodic, twisted, open
 boundaries, infinite system...). The phenomenology of 
 the system and  the  mathematical techniques that are  used
 to analyse it  depend strongly   on the  chosen
 boundary conditions.

 In the present work  we consider the  exclusion process on a finite lattice
   with open boundaries, which  can be viewed as a model for
 a conducting rod in contact with two reservoirs that are not
 in thermodynamic equilibrium with each other (for example,  they are at 
 different temperatures, or have  different chemical, mechanical or electrical
 potentials).  Besides, an external 
 field may be applied to  the system.
 This external field and the  reservoirs drive  a current through the system
 and, in the long time limit,  the connecting rod
 reaches a non-equilibrium stationary state.  
  Our aim is to study
 the statistics of this stationary current. 
  We give explicit formulae, equations~(\ref{GeneralgammaB}) to~(\ref{def:FonctionF}),
  for the cumulant generating function of the   
  totally asymmetric exclusion process (TASEP)   with open boundaries
  that  are valid  for arbitrary
   values of the entrance and exit rates  and for  all values
  of the system size $L$.  We emphasize the fact that 
  our results  are of combinatorial nature 
 and  not only  asymptotic: they 
 describe the TASEP in all  possible regimes, including the
  phase-transition lines. This is in 
  contrast with the asymptotic  expression for the large-deviation function  of the current
  obtained  very recently  using the  Bethe Ansatz \cite{deGierNew}. 
 Indeed, as of today,   the   Bethe Ansatz for the open TASEP
  \cite{deGier1}  seems to be tractable only  in the 
   $L \to \infty$  limit,  and 
    only  inside  the low and high density phases 
  far from the phase-transition  lines.
  We show that the formula obtained  in  \cite{deGierNew} for  $L \to \infty$ 
  can be retrieved
 as a limiting  case of our general  results. The 
 parametric representation we have found  is  obtained by using 
 the Matrix Ansatz technique
  \cite{DEHP,MartinReview}. The  calculations are very cumbersome and 
 some combinatorial patterns   have been guessed  rather than
 fully calculated. The  formulae we obtain must
  therefore be considered as  conjectures.
 But as we shall explain, the results are  exact 
 albeit they have been obtained, at present,  in a  non-rigorous manner.
 They  have  also  been thoroughly  verified 
 against  exact    computations for systems of small sizes. Besides,
 all known special cases can be deduced from our general result. 

 The outline of this work is as follows. In section~\ref{sec:Model}, we recall the 
 definition of the model and the  basic properties that will be used later 
 (the master  equation, the matrix solution and the phase diagram). 
 Section~\ref{sec:resultats}  contains the main results of this work;   we  first restate
 the problem of current fluctuations in the standard mathematical  framework, and
  then  describe
 the simpler  case where all the parameters in the system are equal and taken to be 1.
 The general formulae are presented  in \ref{subsec:general} and are followed by a physical
 discussion of the behaviour of the system through the phase diagram;   we also discuss 
 some  connections  with previously known results.
 In section~\ref{sec:calculs}, we explain  the line  of reasoning
  that we have  followed to obtain
 the formula for the  current fluctuation,  show  where the  unproven assumptions reside,
 and  describe  some  numerical verifications. The last  section is devoted to 
 concluding remarks. A detailed  derivation of the large $L$ limit   \cite{deGierNew} from 
 the general formula is given in the appendix.

\section{Definition of the TASEP  and basic properties}
\label{sec:Model}

\begin{figure}[ht]
\begin{center}
\includegraphics[width=0.70\textwidth]{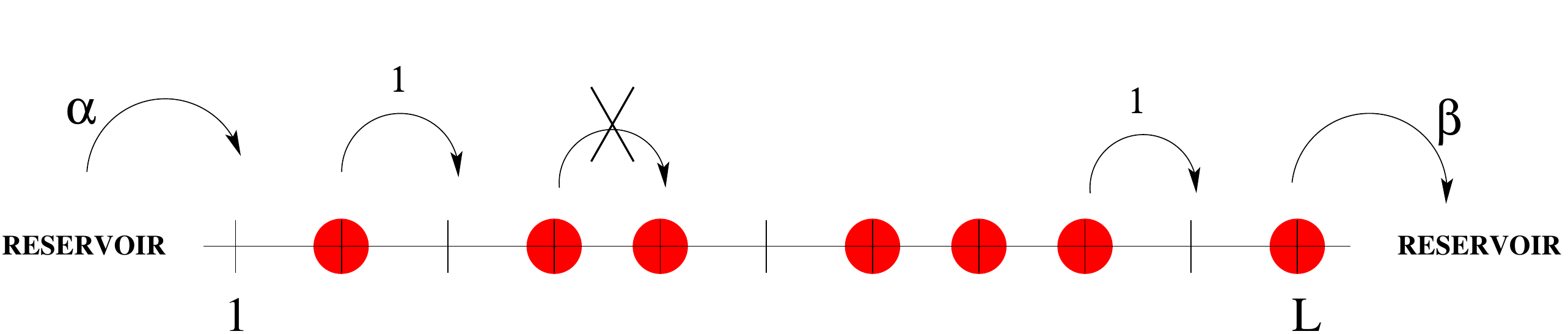}
\caption{Illustration of the TASEP  with open boundaries on a finite lattice with $L$ sites.}
\label{fig-TASEP}
\end{center}
\end{figure}

 We consider the totally asymmetric simple exclusion process (TASEP)
 on a finite lattice of size $L$  with open boundaries. Each site
 of the system can be occupied by at most one particle ({\it exclusion
 condition}). The dynamics of the model is defined by the following
 stochastic  rules (see figure~\ref{fig-TASEP}):  a  particle  at a  site  $i$  in the
 bulk of the system   (with $1 \le i \le L-1$) can jump  with rate $p=1$
 (i.e with  probability  $dt$ during the time interval $dt$)
  to the site $i+1$  if this target site is vacant;
 if site 1 is empty, a particle can enter with rate $\alpha$; 
 a particle  at site $L$ can leave the system with rate $\beta$.
 The entrance and exit rates represent the coupling of the finite
 system with infinite reservoirs  located at its boundaries.
 At a given time, the system is in one of its $2^L$ possible
 configurations and  evolves  according to its stochastic  dynamics
 as  a Markov process. The  evolution of the system 
 can be encoded in the  Markov matrix  $M$ as follows: the probability 
 $P_t({\mathcal C})$ of being in configuration ${\mathcal C}$
 at time $t$ satisfies the master equation 
\begin{equation}
 \frac{d P_t({\mathcal C})} {dt} = \sum_{{\mathcal C'}}
  M({\mathcal C}, {\mathcal C'}) P_t({\mathcal C'}) \, .
 \label{Eq:Markov}
\end{equation}
 The non-diagonal  matrix element $M({\mathcal C}, {\mathcal C'})$ represents
 the transition rate from ${\mathcal C'}$ to ${\mathcal C}$.
 The diagonal part 
  $M({\mathcal C}, {\mathcal C}) = -  \sum_{{\mathcal C'} \neq  {\mathcal C}}
  M({\mathcal C'}, {\mathcal C}) $
  represents (minus) the exit rate from ${\mathcal C}$.
 The Markov matrix is a stochastic matrix: the sum of the elements
 in any given column vanishes.

 In the long time limit, the system reaches a  steady state 
 in which each of the $2^L$ possible configurations occurs with
 a stationary  probability.   This  steady state  probability 
 lies in the kernel of the Markov Matrix: the rules of the ASEP
 ensure that this kernel is non-degenerate and that all 
 other eigenvalues of $M$ have strictly negative real-parts that
 correspond to relaxation states with  a possible  oscillatory behaviour
 (Perron-Frobenius theorem).
 Finding this 
 stationary measure is a non-trivial task: the model is far from equilibrium
 with  a  non-vanishing steady-state  current, 
 there is no underlying Hamiltonian  and  no  temperature.  Therefore
 the fundamental principles of equilibrium statistical mechanics,
 such as the Boltzmann-Gibbs law, cannot be used.

 The   exact calculation of the 
 stationary measure for  the  TASEP  with open boundaries
 and the derivation of its    phase diagram 
 have  played  a  seminal role by 
   triggering  a  whole field of research on exactly solvable models
 in  non-equilibrium statistical mechanics. We recall that the fundamental
 observation  \cite{DeDoMuk} is the existence of recursion relations for the 
  stationary  probabilities between systems of different sizes.
 These recursions are particularly striking when  $\alpha = \beta =1$
 \cite{DeDoMuk}; they can be generalized to arbitrary values
 of  $\alpha$ and  $\beta$  \cite{Schutz, DEHP}
 and also to the more general case  in which
 backward jumps are allowed (PASEP).  The most elegant and efficient way
 to encode these recursions is to use the Matrix Ansatz \cite{DEHP}.
 A  configuration
 ${\mathcal C}$  can be represented by the binary string of length $L$,
 $(\tau_1, \ldots, \tau_L)$,  where $\tau_i = 1$ if the site $i$ is occupied
 and  $\tau_i = 0$ otherwise. Then,    to each 
 ${\mathcal C}$, the following matrix element is associated: 
\begin{equation}
  P(\mathcal C) = \frac{1}{Z_L} \langle \alpha | \prod_{i=1}^L 
 \left( \tau_i {\bf D} + (1 - \tau_i)  {\bf E}\right) | \beta  \rangle \, .
\label{MPA}
\end{equation}
 The scalar  $P(\mathcal C)$,   thus defined, will be equal   to 
 the stationary probability of  ${\mathcal C}$ 
if  the operators $ {\bf D}$ and  $ {\bf E}$, the bra-vector $ \langle \alpha | $
 and the ket-vector $ | \beta  \rangle$ satisfy  the following  algebraic
 relations 
\begin{eqnarray}
             {\bf D} \,  {\bf E} & = & {\bf D} \, + \, {\bf E}    \nonumber \\
             {\bf D}  \,  | \beta  \rangle   & = & 
  \frac{1}{\beta}| \beta  \rangle    \nonumber   \\
              \langle \alpha | \, {\bf E}  & = &  \frac{1}{\alpha} \langle \alpha |  \,\, . 
 \label{DEHPAlgebra}
\end{eqnarray}
 This algebra allows  us to calculate any matrix element of   type~(\ref{MPA}).
 The normalisation constant in equation~(\ref{MPA}) is given by
 \begin{equation}
  Z_L =  \langle \alpha | \left( {\bf D} + {\bf E} \right)^L | \beta  \rangle \, .
\label{def:Z_L}
\end{equation}
 For $\alpha = \beta =1$,   $Z_L$  is a   Catalan number  \cite{DEHP}.  
  More generally, the   Matrix Product Representation method 
  has proved to be very fruitful for  solving   many one-dimensional 
 systems: a very 
 thorough review of this method can be found in \cite{MartinReview}.

\begin{figure}[ht]
\begin{center}
\includegraphics[width=0.35\textwidth]{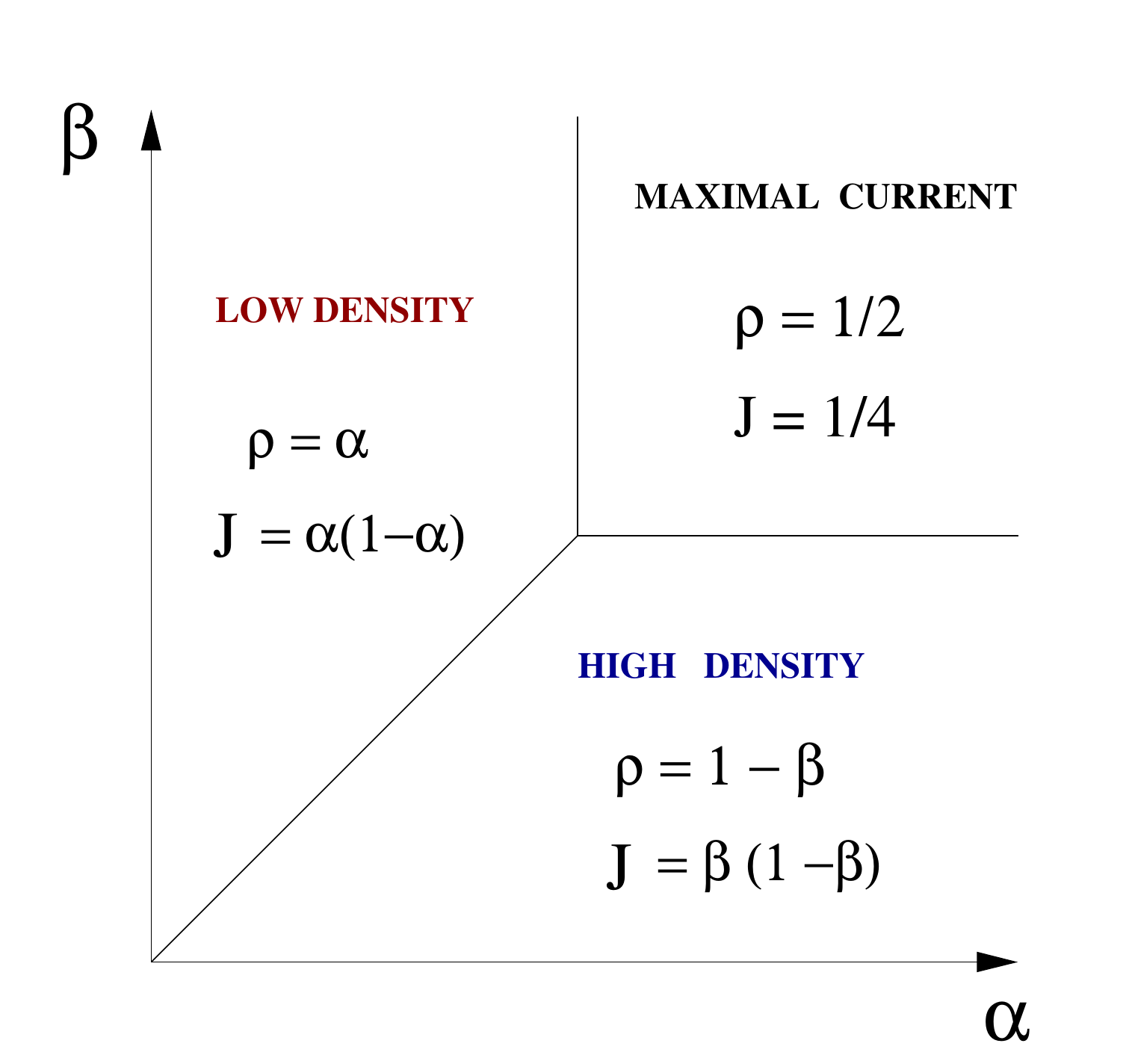}
\caption{The phase diagram of the TASEP  as a function of the boundary rates.}
\label{fig-DIAGPHASE}
\end{center}
\end{figure}

 From the  exact solution, 
 the phase diagram of the TASEP, as well as 
  stationary equal-time correlations and 
 density profiles, 
 can be determined. In the limit of large system sizes,
   the  phase diagram (figure~\ref{fig-DIAGPHASE})  consists of  three main   regions 
\begin{itemize}
  \item For $\alpha < \min(\beta, 1/2), \, $ 
 the system is in the {\it Low-Density Phase}  and its behaviour
 is  driven by the entrance
 rate $\alpha$. The  bulk-density is $\rho = \alpha$
 and the  average current ${J} = \alpha(1 -\alpha)$.  

  \item  The {\it  High Density
 phase}, for  $\beta <  \min(\alpha, 1/2) $ 
 is characterized by  $\rho = 1 - \beta$
 and  ${J} = \beta(1 -\beta)$. 

  \item  In the {\it  Maximal Current Phase,} with  $\alpha >1/2$
 and  $\beta >1/2$, the bulk behaviour is independent of   the
 boundary conditions and one has   $\rho = 1/2$ and 
  ${J} = 1/4$. However,
  correlations with the boundaries  decay  only  algebraically 
 whereas they decay exponentially in the two other phases.

   \item The  low and high density phases are separated by the
 {\it  `shock-line',}  $\alpha = \beta \le 1/2$, across which the  bulk-density
 is discontinuous. In fact, the profile on this  line is a mixed-state
 of shock-profiles interpolating between the  lower density 
 $\rho_l = \alpha$ and the higher density  $\rho_H = 1 - \beta$. 
\end{itemize}

  Detailed properties of the  phase diagram are reviewed for example in 
\cite{DerridaRep,Schutzrev, MartinReview}. We note that the phase diagram
 was obtained  in \cite{Krug}  through physical reasoning
 by using a  hydrodynamic limit and mean-field arguments (see also  \cite{Spohn}).
 However, a  finer analysis does  require the knowledge of the exact solution.

\section{Current fluctuations for the TASEP}
\label{sec:resultats}

       Many  properties of the ASEP have been understood 
 using different  techniques (Matrix Ansatz, Bethe Ansatz, Random Matrices)
 \cite{MartinReview,DerridaRep,FerrariPatrick, 
 OGKM,PaulK,Sasamoto,Schutzrev,Spohn,Zia}.
 However, the determination of current fluctuations in the  original
 TASEP model with open boundaries 
   has remained a vexing and challenging  unsolved  problem. This question 
 is  interesting  and  important: first, 
 in presence of reservoirs, the model is quite realistic
 and can be related to real experimental situations
 \cite{Flindt,VonOppen} (a more detailed discussion can be found
 in \cite{Doucot});  second,
 the exact calculation of fluctuations to all order is akin 
 to determining  the large-deviations of the current which 
 are expected to play a central role 
in non-equilibrium statistical mechanics \cite{DerrReview,Touchette}.

\subsection{Statement of the problem}

    We consider the TASEP with open boundaries and want  to study the
 total (i.e.,  time-integrated) current that has flown through it in the long time limit.
 One way to quantify this total current is to place a counting variable
 $N_t$ at the entrance site. At $t=0$ we have  $N_0 = 0$; each time
 a particle enters the system, we increment the value of  $N_t$ 
 by 1. Hence,  $N_t$  is a random variable that counts the total number of 
 particles that  have entered the TASEP  between time 0 and $t$.
 Because the size $L$ is bounded
 and no particles are created or destroyed in the bulk,  $N_t$
 also represents,  when $ t \to \infty$, 
 the number of particles that have crossed any bond
 in the system, or have exited from  the TASEP from its right boundary.
 We call $N_t$  the total current at time $t$ and we intend to study
 its statistical properties. 

      When $t \to \infty$, the expectation value of  $N_t/t$ converges towards
 the average stationary current  $J$:
\begin{equation}
  \lim_{t \to \infty} \frac{\langle N_t \rangle} {t} = J(\alpha,\beta, L) = 
 \frac{Z_{L-1}}{Z_L} \, . 
\end{equation}
The value of   $Z_L$, defined in equation~(\ref{def:Z_L}),
  was  determined exactly in \cite{DEHP} using  the Matrix Ansatz.

     The variance of  $N_t$ in the long time limit also increases
 linearly with time. This allows to define a `diffusion-constant' $\Delta$,
 as follows:
\begin{equation}
  \lim_{t \to \infty}
 \frac{ \langle N_t^2 \rangle -\langle N_t \rangle^2} {t} = 
 \Delta(\alpha,\beta, L) \,. 
\end{equation}
 An exact expression for  $\Delta(\alpha,\beta, L)$ was obtained
 in \cite{DEMal}. It involved an extension of the matrix method 
 to represent some  two-time correlations in the stationary state.
  For  $\alpha=\beta =1$, the formula for  $\Delta$ is quite elegant and 
 involves  simple  factorial factors (see the next section). Unfortunately,
 the general result is somehow 
 less compact (see equations (58) to (61) in  \cite{DEMal}).

  More generally, one can define moments and cumulants to all orders for the
 random variable  $N_t$. This data  can be encoded in the exponential
 generating function $\langle \exp(\gamma \, N_t) \rangle .$
 The moments of  $N_t$ are obtained by 
 taking  successively the derivatives  of this function at  $\gamma =0$.
 In the long-time limit,  we have 
\begin{equation}
   \langle \exp(\gamma \, N_t) \rangle \simeq \exp(E(\gamma) t ) \, . 
 \label{MomentEigenval}
\end{equation}
(A mathematical equality is obtained by taking the logarithm
 of both sides, dividing by $t$ and taking $t \to \infty$).
 The derivation of this result is recalled in Section~\ref{sec:calculs}. 
 Because  the logarithm of  $\langle \exp(\gamma \, N_t) \rangle $
 generates the cumulants of   $N_t$,  we observe that these cumulants 
 grow linearly with time and that their values are given
 by the derivatives of  $E(\gamma)$  at  $\gamma =0$. The cumulant
 generating function 
 $E(\gamma)$  is  related to the large-deviation function
 of the current by Laplace transform. For finite-size systems, both
 functions carry the same information. 
 
     The function  $E(\gamma)$ was calculated for
  the symmetric exclusion process  in \cite{Doucot}. 
  Very recently, the analysis of the  Bethe Ansatz equations, for  $L \to \infty$
  was carried out for the asymmetric case,  in the 
 low and in the  large density phases \cite{deGierNew}. In the present  work, 
  we obtain an  explicit representation
 of  the  generating function of  the cumulants  of  the  current 
 for all  values of $\alpha$ and $\beta$ and  for all values
 of the system size $L$. The technique used is  
 an extension of the matrix method (see section \ref{sec:calculs}).
 
 For pedagogical reasons 
 we shall  first  discuss  the case
 $\alpha = \beta =1$, which  belongs to the maximal current phase. Here, 
 the formulae are explicit,  quite  simple and  appealing. 
 The general case with arbitrary values of  $(\alpha, \beta)$ will be 
 presented in a separate subsection.

\subsection{The cumulant generating function for  $\alpha=\beta =1$}

 Here, we consider the values  $\alpha=\beta =1$. Historically, the TASEP
 with $\alpha=\beta =1$  was the first case  for which the stationary measure was
 determined exactly \cite{DeDoMuk}.
  For the cumulants of the current, these special values
 lead to  simple  mathematical expressions. 

   The  generating function of the cumulants  $E(\gamma)$
 is given  as a function of $\gamma$ through the following
 representation in terms of a parameter $B$:
\begin{eqnarray}
  \gamma &=& - \sum_{k=1}^\infty
 \frac{(2k)!}{k!} \frac{[2k(L+1)]!}{[k(L+1)]! \, [k(L+2)]!}
 \frac{B^k}{2 k}  \label{gammaB}  \,\, , \\
  E  &=& - \sum_{k=1}^\infty
 \frac{(2k)!}{k!} \frac{[2k(L+1)-2]!}{[k(L+1)-1]! \, [k(L+2)-1]!}
 \frac{B^k}{2 k}   \,\,  . \label{EenB}
 \end{eqnarray} 
 By expressing  $B$ in terms of  $\gamma$   in equation~(\ref{gammaB})
 and substituting in equation~(\ref{EenB}) we can  calculate
  $E$ as a function of $\gamma$  to any desired order.
  The coefficients of this expansion  that we denote by
\begin{eqnarray}
 E(\gamma) = E_1 \gamma + \frac{E_2}{2!}  \gamma^2 +  \frac{E_3}{3!}  \gamma^3
 + \ldots,, 
 \end{eqnarray} 
 give the successive cumulants of $N_t$  in the long time limit.
 For example, we obtain 
 \begin{equation}
    \lim_{t \to \infty}\frac{\langle N_t \rangle}{t} = 
   J = E_1 = \frac{L+2}{2(2L+1)} \, . 
\end{equation}
 This expression is identical to that  found in \cite{DEHP}.
 When $L \to \infty$, $J \to 1/4$ as expected in the maximal current phase.
 At the second order, we obtain 
 \begin{equation}
    \lim_{t \to \infty}
 \frac{\langle N_t^2 \rangle - \langle N_t \rangle^2 }{t} = 
   \Delta = E_2 = \frac{3}{2}
 \frac{(4L+1)! [ L! (L+2)!]^2 }{[(2L+1)!]^3 (2L+3)! } \,.
 \label{DiffConst}
\end{equation}
 This is the same formula as the one derived in \cite{DEMal}.
  When $L \to \infty$, we have
  $\Delta \simeq \frac{3 \sqrt{2\pi}}{64} L^{-1/2} $ i.e.,  the
 diffusion constant vanishes in the  maximal current phase.

The third cumulant, known as the skewness, is given by:
 \begin{equation}
 E_3 = 12 \,  \frac{ [ (L+1)!]^2 [(L+2)!]^4}{(2L+1) [(2L+2)!]^3 }\,
 \Big\{ 9 \frac{(L+1)!(L+2)! (4L+2)!(4L+4)!}{(2L+1)! [(2L+2)!]^2  [(2L+4)!]^2}
- 20 \frac{ (6L+4)!}{(3L+2)!(3L+6)!}  \Big\} \, . 
\label{Skew11}
\end{equation}
 This formula was not known before. For a large system, the  skewness
 behaves as 
\begin{equation}
 E_3 \simeq \frac{2187 - 1280 \sqrt{3}}{10368} \, \pi \sim -0.0090978...
 \label{SkewLimit}
\end{equation}
 Carrying on the elimination of $B$, the next few orders can also
 be found explicitly.  It is  found that the $k$-th cumulant scales
 as $\pi (\pi L)^{k/2 - 3/2}$ for $k \ge 2$. We determined  the
 constant  prefactor for the first few values of $k$.

 The fact that the large deviation function is obtained in parametric
 form should not come as a surprise. On the contrary, this structure  seems
 to be  rather common: to our knowledge it appeared first in \cite{DLeb},
 where the exact  large deviation function for the TASEP on a ring
 was calculated by Bethe Ansatz; it also occurred in the case
  of a defect particle on a ring \cite{evans}. The complete solution
 for the current fluctuations of the ASEP on a ring involves a
 tree-structure that is  again  written  parametrically 
 \cite{Sylvain4}. More generally, it was shown in
 \cite{Bodineau, Bodineau1,DerrReview}
 that for a large class of    non-equilibrium diffusive systems, 
 characterized by a linear conductivity and equilibrium
 diffusion coefficient, the large deviation function can be expressed
 through a parametric set of equations.

 We observe that the limiting  value of the  skewness, given in
 equation~(\ref{SkewLimit}), 
  is a finite number. More generally, one can construct  simple ratios of cumulants
 that   have a finite limit when   $L \to \infty$. 
 It should be possible to `measure' such  numbers  in 
 simulations and to test the universality of these ratios  in a manner
 similar to  \cite{Appert}.

\subsection{The general case}
\label{subsec:general}

 We now give the result for the cumulant generating function,
 valid for any system size $L$ and for  arbitrary values of $\alpha$ and $\beta$. The mathematical
 structure is  the same as above:
\begin{eqnarray}
  \gamma &=& - \sum_{k=1}^\infty  C_k(\alpha,\beta)
 \frac{B^k}{2 k} \,\, ,   \label{GeneralgammaB} \\
  E  &=& - \sum_{k=1}^\infty  D_k(\alpha,\beta)
  \frac{B^k}{2 k}   \,\,  .  \label{GeneralEenB}
 \end{eqnarray} 
  The coefficients $C_k$ and  ${D_k}$   of the series 
  are functions of  the rates  $\alpha$ and $\beta$ and of $L$. Their  explicit expressions
  are  given by 
\begin{eqnarray}
 C_k(\alpha,\beta) &=&  \oint_{\{0,a,b\}} \frac{dz}{2i\pi} 
 \frac{F(z)^k}{z}   \label{def:Ck}  \\
\hbox{ and  }  \,\,\,          
   D_k(\alpha,\beta)  &=&   \oint_{\{0,a,b\}} \frac{dz}{2i\pi}
\frac{  F(z)^k}{(1+z)^2}   \label{def:Dk}
\end{eqnarray}
 where the rational function  $F(z)$ is given by  
\begin{eqnarray}
 F(z) =  
 \frac{ - ( 1 + z)^{2L} ( 1 - z^2)^2 }{z^L (1 -az)(z -a)(1 -bz)(z -b)}
 \,\,\,    \hbox{ with  }  a = \frac{ 1 - \alpha}{ \alpha} \,\, 
 \hbox{ and  }  \,\, b =  \frac{ 1 - \beta}{ \beta} \, .
 \label{def:FonctionF}
\end{eqnarray} 
 
 Equations~(\ref{GeneralgammaB}) to~(\ref{def:FonctionF})  provide
 an exact representation  for  the  generating function of the  cumulants
 of the current in the TASEP
 with open boundaries. These equations  are valid for any system size $L$
 and  for  arbitrary values of the parameters $\alpha$ and $\beta$.
 Before we  proceed further, we  explain the meaning of the
 symbol  $ \oint_{\{0,a,b\}}$: it represents the complex integral along
 3 infinitesimal contours that encircle the points 0, $a$ and $b$
 in the complex plane. Equivalently, using the Cauchy formula  
 we could have written  equations~(\ref{def:Ck}) and~(\ref{def:Dk})  as 
\begin{eqnarray}
  C_k(\alpha,\beta) &=& 
\sum_{z_0 =0,a,b} \hbox{Residue}\left[\frac{F(z)^k}{z}, z_0 \right] \,\, ,   \nonumber  \\
   D_k(\alpha,\beta) &=&  \sum_{z_0 =0,a,b}  \hbox{Residue}  
\left[\frac{F(z)^k}{(z+1)^2}, z_0  \right]  \, . \nonumber  
\end{eqnarray}  
  We note that similar integrals have already appeared in closely  related problems
 \cite{Cedric, evans}. Here, the notations implicitly assume  that $0, a$ and $b$ are distinct.
 If two or three of these numbers coincide, one should  take into account 
  the corresponding  residue only once. Hence, 
 for  $\alpha = \beta = 1$, which implies $a = b =0$, only  the residue
  at 0 must be calculated.
 The resulting  explicit expressions for  $C_k(1,1)$ and  $D_k(1,1)$  give 
  the parametric representations~(\ref{gammaB})  and~(\ref{EenB})  that
 were discussed in the previous section.

 By inverting the series~(\ref{GeneralgammaB}) order by order and substituting the result
 in~(\ref{GeneralEenB}), we can derive closed expressions for the first
 few cumulants of the current. For example, the mean-value of the current
 is given by
\begin{equation}
 J = \frac{D_1(\alpha,\beta)}{C_1(\alpha,\beta)} \,. 
\end{equation}
This formula is, of course,  the same as obtained in \cite{DEHP}. We
 emphasize that  $C_1(\alpha,\beta)$
 and $D_1(\alpha,\beta)$    coincide with  $Z_L$ and $Z_{L-1}$, 
respectively, as can be seen  by comparing  equations~(\ref{def:Ck})
 and~(\ref{def:Dk}) for $k =1$ with  equation  (B10) 
 of   Reference  \cite{DEHP} (see remark 
\footnote{To be fully precise,  $C_1(\alpha,\beta)$
 and $D_1(\alpha,\beta)$   differ from  $Z_L$ and $Z_{L-1}$   by  the  normalisation constant 
 $\kappa^2 = (\alpha + \beta -1)/(\alpha \beta)$  that appears in  \cite{DEHP} and  that ensures
   that $\langle \alpha | \beta \rangle = 1$. In our work, this constant $\kappa^2$  has been
 absorbed in the parameter $B$. If we had made the equivalent choice to use the function
   $\kappa^2 F(z)$  instead of  $F(z)$,  then  $C_1(\alpha,\beta)$
 and $D_1(\alpha,\beta)$ would be  identical to  $Z_L$ and $Z_{L-1}$.}).

 At second order we obtain an expression for the diffusion constant
\begin{equation}
 \Delta = \frac{D_1 \, C_2 - D_2\, C_1}{C_1^3} \,.
\label{Deltaabgene} 
\end{equation}
 We remark  that $C_2$ and $D_2$ are natural generalisations of $C_1$ and $D_1$.
 In this form, the diffusion constant
 looks more compact than the formula  found in \cite{DEMal}. The two expressions
 must coincide but we have not endeavored to prove  this fact analytically  for all
 values of $\alpha,\beta$ and $L$.
 We can carry on this procedure further to a few more orders to obtain higher cumulants
 and we did so with the
 help of a symbolic  mathematical calculation tool.

 It is of greater interest to 
 analyse  the  behaviour of the large-deviation function in
 the different phases of the model  when  $L \to \infty$.
\begin{itemize}
\item In the Maximal Current phase, with $\alpha >1/2$ and  $\beta >1/2$, 
 the parameters  $a$ and $b$ lie  inside the unit circle.  Hence, the contour integrals that
 appear in equations~(\ref{def:Ck}) and ~(\ref{def:Dk}) can be replaced by a 
 single  integral along the unit circle. Then, we apply the saddle-point method to estimate
 the asymptotic behaviour of $C_k$ and $D_k$ when $L \to \infty$.
 The saddle point is at $z=1$ and we  observe that
 the values of $a$ and $b$ do not influence the saddle-point
  estimation at dominant order: in fact they
  contribute by   the same multiplicative factor  $[(1-a)(1-b)]^{2k}$, 
  that can be reabsorbed in the parameter $B$. Therefore, the behaviour of the cumulants in 
  the large $L$ limit does not depend on the boundary rates $\alpha$ and $\beta$ in the 
  Maximal Current phase (as expected).
 The results  at dominant order  are the same as those  obtained for $\alpha=\beta =1$, the special
 case  discussed in the previous section.

\item In the Low Density  phase,  $\alpha < \min(\beta, 1/2)$,  the parameter  $a$
 is outside the unit circle, we have  $a > b$ and
 the position of $b$  with respect to the unit circle  is not determined.  In the large
 $L$ limit, it is the pole at $z =a$ that contributes dominantly to the values
 of $C_k$ and $D_k$ and the parametric representation~(\ref{GeneralgammaB})
 and~(\ref{GeneralEenB})  becomes  (see the appendix for more details)
\begin{eqnarray}
 \gamma &=&  -\sum_{k=1}^\infty B^k \frac{1}{k!}
 \frac{d^{k-1}}{dz^{k-1}} \Big \{  \frac{\phi^k(z)}{z} \Big\}  \Big|_{z =a} 
\label{GversusBLD}\\
   E  &=&  -  \sum_{k=1}^\infty
B^k \frac{1}{k!}
 \frac{d^{k-1}}{dz^{k-1}} \Big\{  \frac{\phi^k(z)}{(z+1)^2}  \Big\} \Big|_{z =a} \, ,
\label{EversusBLD}
 \end{eqnarray} 
 where  the  function  $\phi(z)$ is given by  
\begin{eqnarray}
\phi(z) =  (z -a) F(z) =  
 \frac{ - ( 1 + z)^{2L} ( 1 - z^2)^2 }{z^L (1 -az)(1 -bz)(z -b)} \, .
 \label{def:FonctionPHI}
\end{eqnarray}
 These expressions can be used to calculate the first few cumulants  in the low density phase:
\begin{eqnarray}
         E_1 &=&   \rho ( 1 -  \rho) \, \nonumber \\
         E_2 &=&   \rho ( 1 -  \rho)( 1 - 2 \rho) \, \nonumber \\
         E_3 &=&   \rho ( 1 -  \rho)( 1 - 6 \rho + 6 \rho^2 )  \, \nonumber \\
         E_4 &=&  \rho ( 1 - \rho)( 1 - 2\rho)( 1 - 12\rho + 12\rho^2 )   \nonumber \\
           E_5 &=& \rho ( 1 -  \rho)( 1 - 30\rho + 150 \rho^2 -240 \rho^3
   + 120 \rho^4)  \,\,  \hbox{etc...}
\end{eqnarray}
 with the mean density $\rho = \alpha$.

  In fact, using  equations~(\ref{GversusBLD}) and~(\ref{EversusBLD}), the function 
 for $E(\gamma)$   can be obtained in a closed form thanks to the Lagrange
 Inversion Formula \cite{Flajolet}, as explained in the appendix. This leads  to 
\begin{eqnarray}
    E(\gamma) = \frac{a}{a+1} \frac{ {\rm e}^\gamma -1 }{{\rm e}^\gamma + a} \,.
\label{LimitLD}
\end{eqnarray}
 This expression  is identical in the  TASEP case
   to  the one  obtained by Bethe Ansatz in \cite{deGierNew}.
 (In \cite{deGierNew}, the general PASEP is studied: this  adds  a prefactor $(p -q)$
 to~(\ref{LimitLD}), where $p$ and $q$ are  the rates of forward
 and backward jumps, respectively, 
 and modifies the definition of $a$).
 We remark that the limit formula~(\ref{LimitLD}) is rather simple and that  it is  
   {\it totally  independent} of  the specific  form of the 
  function  $\phi(z)$, as can be seen from  the derivation given in the appendix. 
 An elementary and  physical 
 derivation of this result can be given by using macroscopic fluctuation theory 
 \cite{BoDerr, DerrPriv}.

\item The High Density phase is symmetrical to  the low density phase under the exchange of 
 $\alpha$ and  $\beta$. Therefore   a separate discussion is not  required.

\item The shock line $\alpha = \beta \le 1/2$, i.e.,  $a = b \ge 1$ can also be analyzed
 from our  general formula. We have calculated explicitly the first few cumulants and 
 obtained the following scaling behaviour
 $$  E_k  \simeq   \epsilon_k     \alpha(1 - \alpha) ( 1 - 2 \alpha)^{k -1} L^{k-2}  
 \hbox{ for } \, k \ge 2\, .$$
 We recall that the current is given by  $E_1 = \alpha(1 - \alpha)$. The   numerical
 coefficients are given by  $\epsilon_2 = 2/3$,   $\epsilon_3 = -1/30$, 
 $\epsilon_4 = 2/315$,  $\epsilon_5 = -1/1890$... The fact that the higher cumulants
  grow without bounds with $L$ whereas they are bounded in low and high density phases
  may come as a surprise. However, it is known  that in the  shock phase, particles
  have a vanishing chemical potential and the equivalence between canonical and grand-canonical 
  descriptions  breaks down \cite{km,Speer}.

 More precisely, this puzzling scaling  can be understood  from
   the domain-wall picture,  introduced in \cite{DEMal}  to describe 
  the discontinuity  by a factor $2/3$ in  the diffusion-constant
 along the shock line. In the limiting case $\alpha = \beta \ll 1$, the dominant configurations
 are shock profiles  between a  low density  region with  $\rho_{l} = \alpha$
 and a high density  region  with  $\rho_{h} = 1 - \beta$. The  shock
 is localized on a single site.  The dynamics  becomes equivalent to that of a symmetric
 random walker on a lattice of $L+1$ sites and confined by
  two reflecting boundaries. The number of particles having entered
 the TASEP corresponds  to the number of leftward steps of the shock. 
 The statistics  of  the  steps performed
 by an effective  random walker between  two reflecting walls is a well-posed problem that
   can be studied  independently. For instance, 
   we find that the skewness (third cumulant) of the  random walker grows as $-L/30$.
 However, if we introduce some bias in the jumping rates (which corresponds to  the high
 or low density phase)
  the walker becomes localized near
 one of the walls and all cumulants  remain finite when $L \to \infty$. Finally, 
 the  $\epsilon_k$'s  can be calculated  by remarking that the 
  random walker between  two reflecting walls can be mapped to the TASEP  with 2 identical  particles
  on a periodic lattice, and then  using the results of  \cite{DLeb}. 

\end{itemize}

\section{Outline  of  the Analytical Procedure}
\label{sec:calculs}

 We now describe   briefly how the expressions~(\ref{GeneralgammaB}) to~(\ref{def:FonctionF}) 
  were obtained. Detailed explanations are deferred
 to a forthcoming article. 

\subsection{General setup}

 In order to study the statistics of  $N_t$, the total  number of particles that have entered
 into the system, one introduces  the joint  distribution
 $P_t({\mathcal C}, N)$, the  probability of being at time $t$ in configuration ${\mathcal C}$
 and having  $N_t=N$. The evolution equation of $P_t({\mathcal C}, N)$  can be written using 
 the Markov equation~(\ref{Eq:Markov}). It is then useful to introduce the Laplace transform
 \cite{DLeb,OGKM}
 $$F_t({\mathcal C}) = \sum_N {\rm e}^{\gamma N}  P_t({\mathcal C}, N) \, .$$
 These generalized weights satisfy a deformed Markov equation   
\begin{equation}
 \frac{d F_t} {dt} =   M(\gamma)  F_t  \, . 
 \label{Eq:MarkovDef}
\end{equation}
The matrix  $M(\gamma)$  of size $2^L$ is given by
\begin{equation}
  M(\gamma) = M + ({\rm e}^\gamma - 1)M_1 \, , 
\end{equation}
 where $M$ is the original Markov operator and  the matrix $M_1$ 
 contains only  those  transitions in which  a particle enters the system,
 i.e.,  $M_1({\mathcal C}, {\mathcal C'}) =\alpha  $ if  ${\mathcal C'}$ evolves into
  ${\mathcal C}$ by adding a particle at site 1 and  $M_1({\mathcal C}, {\mathcal C'}) = 0$
 otherwise.  Equation~(\ref{Eq:MarkovDef}) is formally solved as  $F_t = \exp(M(\gamma) t) F_0$
  which,  in  the long time limit, behaves as
  $$F_t  \simeq {\rm e}^{E(\gamma) t} \,  |E(\gamma)\rangle \, , $$
 where  $E(\gamma)$ is the dominant   eigenvalue of  $M(\gamma)$ and 
  $|E(\gamma)\rangle$  the associated eigenvector
  (it is unique for sufficiently small  $\gamma$).  Thus, 
 we obtain  equation~(\ref{MomentEigenval})
 $$ \langle \exp(\gamma \, N_t) \rangle =  \sum_{{\mathcal C}} F_t({\mathcal C}) 
 \simeq \exp(E(\gamma) t ) \, .$$
 The cumulant generating function is  therefore identical 
 to the largest eigenvalue of the deformed operator  $M(\gamma)$
 and the problem  of  determining 
 the statistics of  $N_t$ has been traded for  a spectral  problem.
 This question can be tackled using different
 methods. Bethe Ansatz is one technique for integrable systems. Another approach, valid
 for small values of $\gamma$,  is to perform a  perturbative expansion around the dominant
 eigenvector, $|0\rangle$,  and the dominant eigenvalue, $0$, of the original Markov matrix $M$:
\begin{eqnarray}
  E(\gamma) &=& E_1 \gamma + \frac{E_2}{2!}  \gamma^2 +  \frac{E_3}{3!}  \gamma^3
 + \ldots   \nonumber \\
  |E(\gamma)\rangle  &=& |0\rangle + \gamma  |1\rangle + \gamma^2 |2\rangle   + \ldots
\end{eqnarray}
  Carrying out the perturbative expansion explicitly, we find that the  $k$-th order
  correction to the  eigenvector satisfies an equation of the type
\begin{equation}
   M |k\rangle  = {\mathcal R}_k\left( |0\rangle, \ldots  |k-1\rangle  \right) \, ,
\label{eq:ordrek}
\end{equation}
 ${\mathcal R}_k$ being a linear functional.  For example,  we have
$$ M |0\rangle  = 0\, , \,\,\,\,  M |1\rangle  = (E_1 - M_1) |0\rangle \quad   \hbox{etc ...}$$
 Moreover, the solvability condition (obtained by using the fact that  
 vector $\langle 0| = (1,1,\ldots,1)$ 
 is the  left null-eigenvector of the Markov matrix $M$) allows to express
 the  $k$-th  term in $E(\gamma)$ (i.e.,  the cumulant of order $k$) as a linear function 
 of the  $k-1$ vectors $|1\rangle,  |2\rangle, \ldots  |k-1\rangle$. For example,  we have
\begin{equation}
  E_1 = \langle 0| M_1 |0\rangle \,, \,\,\, \,\, E_2 = 2 \langle 0| M_1 |1\rangle
  + ( 1 - 2 E_1)  \langle 0| 1 \rangle  \quad  \hbox{etc...}
\label{RelationsEk}
 \end{equation}

 The  cumulants of the current can be found if we are able to solve the set of linear
 equations~(\ref{eq:ordrek}) for all values of $k \ge 0$.
 At  order 0, the stationary state $| 0 \rangle$ and the mean current  $E_1$ were
 calculated thanks to the Matrix Ansatz,  recalled  in equations ~(\ref{MPA})~(\ref{DEHPAlgebra})
 and (\ref{def:Z_L}).  In \cite{DEMal}, the first order correction 
  was also obtained by a  Matrix Ansatz: the required algebra 
 was constructed by taking tensor products of two quadratic algebras
 of the type~(\ref{DEHPAlgebra}). This allowed  to  calculate the  diffusion constant $E_2$.

 \subsection{Steps of the calculations and Numerical tests}

  The computations that we have carried out can be summarized by the
 following steps: 

\hfill\break 
  (i) The
  Matrix  Ansatz used in \cite{DEMal} has been simplified  and   generalized 
  to all orders. The operators required  to calculate the  $(k+1)$-th cumulant are 
 denoted by  ${\bf D}_k$ and $ {\bf E}_k$. These operators   are constructed  by using  $2k+1$
 tensor products of the original ${\bf D}$ and ${\bf E}$'s. This may appear 
 daunting at first sight but in fact this
 is not so bad: the  ${\bf D}_k$ and  ${\bf E}_k$ already 
 appear in the studies of   multispecies exclusion processes  \cite{EFM,MMR,Sylvain4}. There
 they were introduced  as formal objects to construct the stationary measure.
 Here, they are used  as  tools for calculations. We have also found  the  boundary vectors
 ${\bf \langle \alpha_k|}$ and  ${\bf | \beta_k\rangle}$.

\hfill\break 
 (ii) The fact   that  ${\bf D}_k$ and  ${\bf E}_k$ allow to solve
 the system~(\ref{eq:ordrek}) at each order has been  {\it proved recursively}. 
 Using the relations~(\ref{RelationsEk}), 
 an expression for the cumulant  $E_k$ in terms of 
  matrix elements involving ${\bf D}_k$ and  ${\bf E}_k$ and the
 previously determined  $E_j$'s for $j < k$ can be   written. 

\hfill\break  
 (iii) The next step  is to calculate 
 the matrix elements. Typically, one has to determine a `normalization'-type term
  $ \langle \alpha_k| ({\bf D}_k + {\bf E}_k)^L | \beta_k\rangle$ that generalizes~(\ref{def:Z_L}).
 Such a matrix element  can be  found by using  the image method
 \cite{DEMal,MartinNotes} in a space of dimension $2k+1$.
  The resulting expression  involves sums of products
 of binomial factors.  These binomial factors can be  expressed  as multiple contour integrals and the
 total sum can be recast    as a determinant. As this stage, the cumulants are written as
  complex integrals over determinantal  expressions. 
 
 \hfill\break  
 (iv) The diffusion constant  ($k =2$) was recalculated.  
 The skewness ($k =3$) was then  determined by  evaluating the 
 contour integrals.  These integrals  produce  a `generic' term
 and a  very large number of  boundary terms. It was found that the global 
 contribution of the boundary terms cancels  out for $k=2$ and 3. It was also observed that a parametric
 expression for $\gamma$ and $E$  at order 3 as a function of an arbitrary parameter $B$  
 allows to retrieve the diffusion constant and the skewness. 

\hfill\break  
 (v) At  $k$-th order, a  calculation of the `generic' term  was performed {\it  by assuming  that 
 the various  contributions  of boundary terms  globally cancel out}.  It was  observed 
 that the  resulting formula
 could  also be obtained through the   parametric
 expressions~(\ref{GeneralgammaB}) and~(\ref{GeneralEenB})
  for $\gamma$ and $E$ at order $k$ with respect to $B$, that were  {\it   guessed}  at all orders.

\hfill\break  

 We admit that our  final result was guessed rather than fully
 worked out. The gaps in our derivations (steps (iv) and (v)) have to 
 be filled. However, we emphasize that we have been using 
 a systematic procedure and that we are absolutely sure that the 
  equations~(\ref{GeneralgammaB}) to~(\ref{def:FonctionF}) 
  are correct.   We have tested numerically the  conjectured  formulae
 in the following cases:
\begin{itemize}
\item For $\alpha = \beta =1$, the formula~(\ref{Skew11}) for the skewness
 was compared with the  exact result obtained by solving  the linear equations~(\ref{eq:ordrek})
 for the first three orders for systems of  sizes $L \le 10$. The results agreed in  all  cases.
 The expressions are rational numbers which involve integers having more than 20 digits.

\item For $\alpha = \beta =1$, the first 6  cumulants were verified for $L \le 10$.

\item The formula for the second cumulant~(\ref{Deltaabgene}) as a function of 
  $\alpha$ and $ \beta$ was tested against the exact result for $L=4$.

\item We have chosen arbitrary values of $\alpha$ and $ \beta$ throughout  the phase diagram
 (a dozen of different  cases). By taking rational values for  $\alpha$ and $ \beta$, we insure that
 the formal mathematical program (Mathematica) performs exact calculations on integers when solving
  the linear equations~(\ref{eq:ordrek}). 
 We have worked with systems of  size $L \le  10$  and tested the formula up to the 6th cumulant.
 The integers that appear involve hundreds of digits (450 in the worst chosen case): the 
expressions~(\ref{GeneralgammaB}) to~(\ref{def:FonctionF}) give  the correct answer  in all  cases.

\end{itemize}

 Finally, we  emphasize the formulae given here 
 allow us  to re-derive results that were  known  previously, such as  equations~(\ref{DiffConst})
 and~(\ref{LimitLD}). This   provides   an indirect verification
  of their correctness.

\section{Discussion}

 In the present work,
 we give  closed  formulae for the  generating function of the cumulants of the 
 current  in the open TASEP. These results hold  for all  values of the 
  boundary parameters $\alpha$ and $\beta$ and all values
 of the system size $L$.  We emphasize the fact that  our expressions  are exact
 and  not only  asymptotic: they are of combinatorial nature. They allow us  to 
 describe the ASEP  at  all  points of  its phase diagram, including the
phase-transition lines.  The  cumulant  generating function is given 
  in the form of  a parametric representation
 -- equations ~(\ref{GeneralgammaB}) to~(\ref{def:FonctionF}): 
  a  similar   mathematical structure 
   can be found in  other works  \cite{DLeb,evans,Sylvain4} and it  can be related   physically
  to  the additivity principle \cite{Bodineau,DerrReview}.

  The statistics of the current in the open  TASEP had remained a challenging open problem
 for many years and no exact solution for the full  distribution of the current was known
 for finite size systems. The Bethe Ansatz equations for the open ASEP were studied in  \cite{deGier1}
 but they are   valid  only on  some surfaces 
  in the parameter space:  this restriction seemed  to be a major  obstruction to   the computation
 of the large-deviation function. Only 
 recently,  a  very subtle  analysis of the  Bethe  equations,  valid in the   $L \to \infty$
 limit, together with some conjectures on  the asymptotic locations  of the Bethe roots,
 was carried out  in \cite{deGierNew}, leading
  to an expression for  the  cumulant  generating function.
  However, the  result in \cite{deGierNew}  can only be established 
 deep  inside  the low and the high density phases and it is an asymptotic expression, 
 valid only when    $L \to \infty$. 
 In our work,  we have followed a different path and 
 used  the Matrix Product Representation \cite{DEHP,MartinReview} to calculate 
 the cumulants of the current  order by order. We have performed some explicit calculations
 and uncovered the general structure of the solution.  From    the mathematical
 point of view,  our formulae are     only   a  conjecture  but we have verified  it in  dozens
 of cases and derived from it all the  previously known results.  We have absolutely  no doubt that
 our expressions  are true.    In particular, the 
 main expression obtained in 
 \cite{deGierNew}  
 can be derived  as a limiting case of our results.

   We have decided to present the final formulae  before completing
   the proof,  because we find
   the results   elegant and   sufficient by themselves. Furthermore,  
  they allow to draw some
  interesting physical consequences and to  open  new  problems. 
  We think that   the proof  is  a question of carrying out 
   a very  long  computation  rather than having  some deep mathematical insight.
  We are presently working on this aspect. We hope to have given 
  enough details to  the reader to  clarify   what kind of 
  assumptions  were  made,  that
  allowed us to jump to the final result and to guess the full structure of the solution.
   Another possibility,  
  now  that the final formula is  known, is to search for  a direct method to check it:
   after all,  the  cumulant  generating function is nothing
   but the largest eigenvalue of a known operator.

  Besides  completing   the derivation, we also intend to extract from
 equations~(\ref{GeneralgammaB}) to~(\ref{def:FonctionF})   the scaling limit of
 the large deviation function in the  $L \to \infty$ limit. It would
 be interesting to compare that scaling form 
  with  recent numerical results obtained 
 by Monte-Carlo and  DMRG methods \cite{Gorissen,Mitsudo,Proeme}
 and also to  investigate the crossover with theoretical
 results derived  on the
 infinite lattice \cite{FerrariPatrick,Sasamoto}.

 Finally, we have  considered here only  the TASEP. But the matrix method can also be applied
 to the partially asymmetric case (PASEP), and we have checked that tensor products
 of the PASEP quadratic algebra  that appeared in multispecies PASEP models \cite{PEM}
 allow us to solve  the hierarchy of equations~(\ref{eq:ordrek}) for the cumulants. We believe
 that the  parametric representation still holds in the PASEP case and that  combinatorial
 tree-structures akin to those found for the  PASEP on a periodic ring in  \cite{Sylvain4}
 will  probably  play an important role. Besides,  complex integral
 representations for matrix elements
 analogous to those  we have used here also appear in  the PASEP 
\cite{MartinReview,MartinPASEP,SasaPasep1,SasaPasep2}. There
 must exist a general and hopefully elegant  structure that encompasses
 all the cases, though we  
   are well  aware that such a structure  may be difficult to discover.

\hfill\break
 We acknowledge  interesting discussions with  O. Golinelli,
 S. Prolhac  and R. Vasseur.
 We are grateful to  S. Mallick for a very careful reading of the manuscript
 and P. Krapivsky for useful remarks.  We are  thankful
 to B. Derrida for telling us how to  obtain
  equation~(\ref{LimitLD})  from  macroscopic fluctuation theory.

\appendix

  \section{Derivation of equation~(\ref{LimitLD})}

 Let  $C_0$,$C_a$ and $C_b$  be three infinitesimal contours (circles) that
 surround the points $0$, $a$ and $b$ in the complex plane.
 According to equations~(\ref{def:Ck})
 and~(\ref{def:Dk}), we must calculate contour integrals along these three circles.
 We denote by  $\Gamma_1$ the unit circle.

\begin{figure}[ht]
\begin{center}
\includegraphics[width=0.475\textwidth]{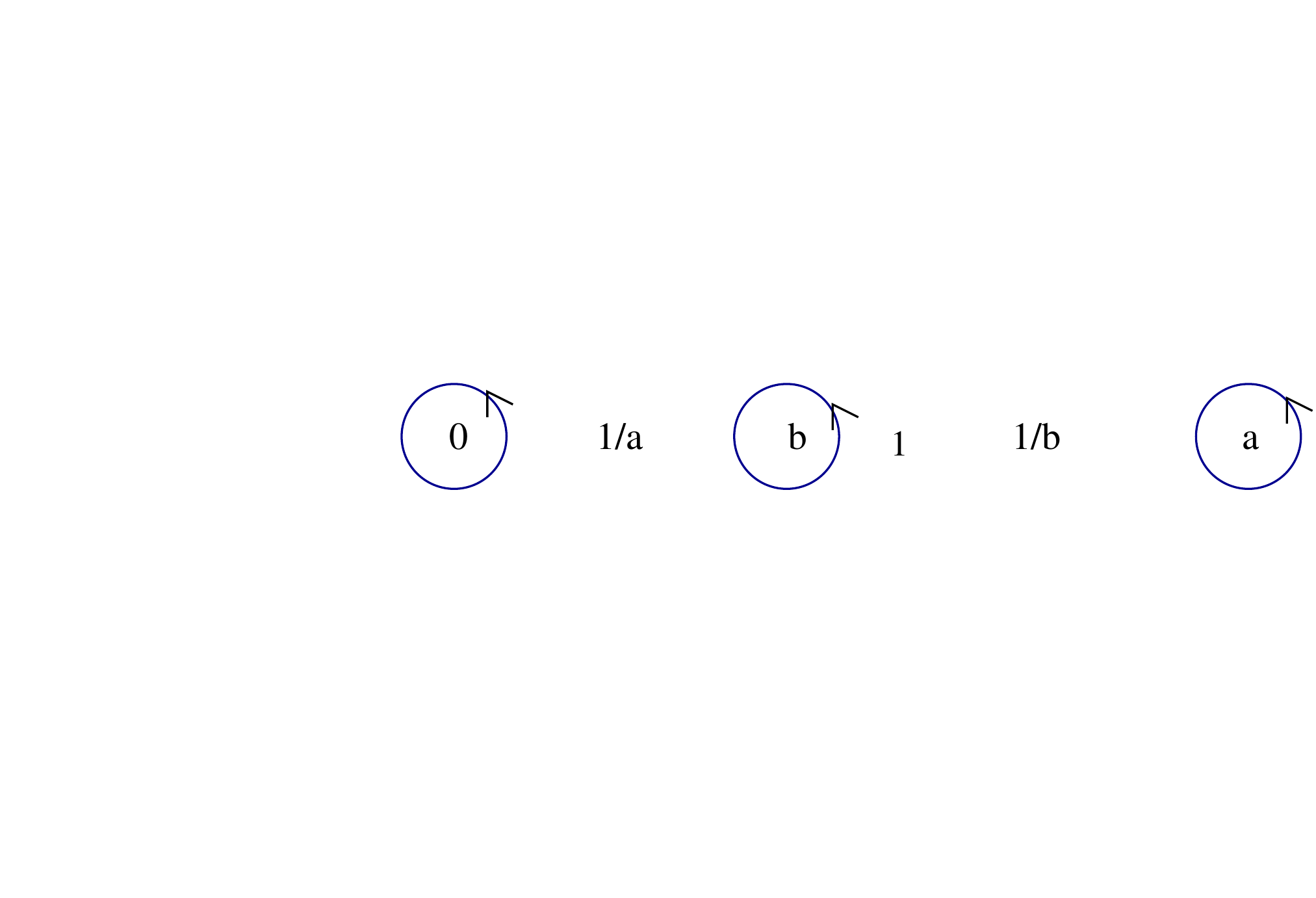}
 \hskip 0.75cm 
\includegraphics[width=0.475\textwidth]{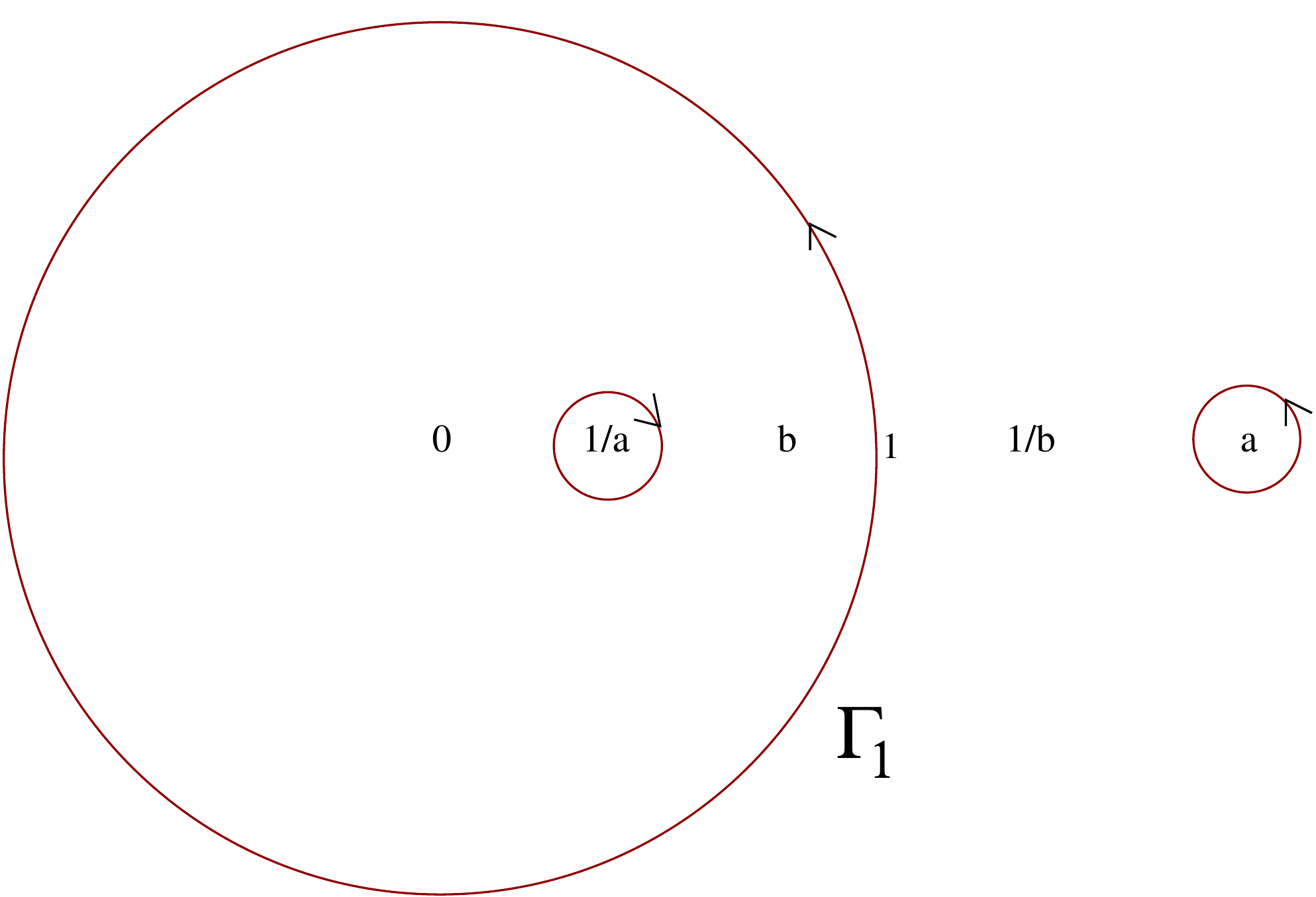}
\caption{The two sets of  equivalent  contours in the complex plane. In this
 figure, we have supposed that $b$ lies  inside the unit disk.}
\label{fig-CONTOURS}
\end{center}
\end{figure}

 In the low density phase $\alpha < \min(\beta, 1/2)$,  the parameter  $a$
 is outside the unit disk  and  we have  $a > b$. The point $b$ can be either
 inside or outside the  unit disk. In order to perform a saddle-point analysis
 in the large $L$ limit,  a  contour must  pass through the point $z=1$.  We therefore deform the
  infinitesimal contour $C_0$ encircling 0 into the unit circle  $\Gamma_1$.  But then we must
 remove the contribution of the pole $1/a$ of $F(z)$ which
  is inside  $\Gamma_1$ (see figure~\ref{fig-CONTOURS}).
 We must also distinguish the cases $ b < 1$ or $b>1$:
 For $b < 1$ we can write, formally: $C_0  + C_a + C_b = \Gamma_1 - C_{1/a} +  C_a$.
 For  $b > 1$ we can write  $C_0  + C_a + C_b = \Gamma_1 - C_{1/a} - C_{1/b}   +  C_a +  C_b$.
 Now the function $F$ satisfies: $F(z) = F(1/z)$. Thus for $z_0 =a,b$, we have 
\begin{eqnarray}
  \hbox{Residue}\left[\frac{F(z)^k}{z}, z_0  \right]   &=&
 -  \hbox{Residue}\left[\frac{F(z)^k}{z}, 1/z_0  \right]  
 \nonumber \\ 
   \hbox{Residue}\left[\frac{F(z)^k}{(z+1)^2}, z_0  \right] &=&  - 
\hbox{Residue}\left[\frac{F(z)^k}{(z+1)^2}, 1/z_0  \right]  \, . 
\end{eqnarray}
Formally this can be  written as  $C_{1/a} = -  C_a$ and  $C_{1/b} = -  C_b$. Finally the contours
 for the complex  integrals reduce to  $\Gamma_1  + 2  C_a$ if $b < 1$ and to 
 $\Gamma_1  + 2  C_a + 2 C_b$ if $b >  1$.

 The integral over  $\Gamma_1$ is estimated by saddle point: it is of order $4^L$.
 The  integral over  $ C_a$  is of order $(2 +  a + 1/a)^L \gg 4^L$ (because $a >1$).
 The residue at $b$  contributes only when $b >1$ but in that case it is 
 of order $(2 +  b + 1/b)^L \ll (2 +  a + 1/a)^L$ (because  $b < a$).
 Hence, in all cases, in  the large $L$ limit, the dominant
 contribution comes from the contour around $a$, 
 which has to be counted twice.
 Finally, using Cauchy's formula, we can write
\begin{equation}
 C_k(\alpha,\beta) = 2 \oint_{C_a} \frac{dz}{2i\pi} \frac{(\phi(z)^k/z)}{(z -a)^k}
  = \frac{2}{(k-1)!} \frac{d^{k-1}}{dz^{k-1}}\left\{  \left(\phi(z)\right)^{k} \frac{1}{z}
  \right\}\Big|_{z =a} 
\end{equation}
 where the function $\phi(z)$ has been defined in~(\ref{def:FonctionPHI}).
Similarly, we have
\begin{equation}
 D_k(\alpha,\beta)
  = \frac{2}{(k-1)!} \frac{d^{k-1}}{dz^{k-1}}\left\{  \left(\phi(z)\right)^{k} \frac{1}{(z+1)^2}
 \right\}\Big|_{z =a} 
 \, .
\end{equation}
 Substituting these expressions into
 the formulae~(\ref{GeneralgammaB}) and~(\ref{GeneralEenB}) leads to 
equations~(\ref{GversusBLD})  and~(\ref{EversusBLD}).

  We  now  need  the Lagrange Inversion formula \cite{Flajolet},
 that  can be stated as follows. Suppose
 that the two complex variables $z$ and $w$  are related in terms
 of the parameter $B$ as
\begin{equation}
 w = z + B \phi(w)
\end{equation}
 where $\phi$ is locally analytic. Then, any function $G(w)$ can be expanded
 as a power-series in $B$ as follows:
 \begin{equation}
   G(w) = G(z) + \frac{B}{1!} \phi(z) G'(z) +  \frac{B^2}{2!}
\frac{d}{dz}\left\{  \left(\phi(z)\right)^2 G'(z) \right\}+ \ldots  +
\frac{B^{n+1}}{(n+1)!}
\frac{d^n}{dz^n}\left\{  \left(\phi(z)\right)^{n+1} G'(z) \right\}+  \ldots 
\label{Lagrange}
 \end{equation}

  We can  identify  the expression~(\ref{GversusBLD})  for $\gamma$
 with the  Lagrange's Inversion  formula
  applied  at the point  $z =a$  with $G(z) = -\log(z)$. We obtain
\begin{equation}
 \gamma =  -\log(w) +  \log(a) \, .
 \end{equation}

 We can also  compare  Lagrange's Inversion  formula with
  the expression~(\ref{EversusBLD}) for $E$, taking now $G(z) = \frac{1}{z+1}$. We obtain
 \begin{equation}
 E = \frac{1}{w+1} -  \frac{1}{a+1} \,.
 \end{equation}

 Eliminating $w$ between the last two equations 
 leads to the desired expression~(\ref{LimitLD}).
 We note that the precise form  of $\phi(z)$ has played no role. This suggest that 
 the formula~(\ref{LimitLD}) for the cumulant generating function should be rather
 universal.

\end{document}